\documentclass[10pt]{article}
\usepackage{graphicx}
\usepackage{hyperref}

\def\Title#1{\begin{center} {\Large #1 } \end{center}}
\def\Author#1{\begin{center}{ \sc #1} \end{center}}
\def\Address#1{\begin{center}{ \it #1} \end{center}}

\newcommand\pubblock{\rightline{\begin{tabular}{l} Proceedings of the Second Annual LHCP\\
         \pubdate  \end{tabular}}}

\newenvironment{Abstract}{\begin{quotation} \begin{center} 
             \large ABSTRACT \end{center}\bigskip 
      \begin{center}\begin{large}}{\end{large}\end{center} \end{quotation}}

\newenvironment{Presented}{\begin{quotation} \begin{center} 
             PRESENTED AT\end{center}\bigskip 
      \begin{center}\begin{large}}{\end{large}\end{center} \end{quotation}}

\def\Acknowledgements{\bigskip  \bigskip \begin{center} \begin{large}
             \bf ACKNOWLEDGEMENTS \end{large}\end{center}}

\def\beq{\begin{equation}}
\def\eeq#1{\label{#1}\end{equation}}
\def\eeqn{\end{equation}}

\def\beqa{\begin{eqnarray}}
\def\eeqa#1{\label{#1}\end{eqnarray}}
\def\eeqan{\end{eqnarray}}

\let\bar=\overbar

\def\ie{{\it i.e.}}

\def\Dslash{\not{\hbox{\kern-4pt $D$}}}
\def\dslash{\not{\hbox{\kern-2pt $\del$}}}

\def\msb{{\bar{\ssstyle M \kern -1pt S}}}

\usepackage{color}

\newcommand{\madanalysis}{{\sc MadAnalysis~5}}

\newcommand{\delphes}{{\sc Delphes}}

\newcommand{\cpp}{{\sc C++}}

\def\ma{\madanalysis}

\newcommand\pubdate{\today}

\def\affiliation{
  LPSC, Universit\'e Grenoble-Alpes, CNRS/IN2P3, \\
  53 Avenue des Martyrs, F-38026 Grenoble, France
 }

\def\support{\footnote{The work presented here was
supported in part by the French ANR projects 12-BS05-0006 DMAstroLHC
and 12-JS05-002-01 BATS@LHC, the Theory-LHC-France initiative of
CNRS/IN2P3,
and the ``Investissements d'avenir, Labex ENIGMASS''. }}

\begin{document}

\large
\begin{titlepage}
\pubblock

\vfill
\Title{  
Implementation and validation of the LHC SUSY searches with MadAnalysis 5 }
\vfill

\Author{ B\'eranger Dumont \support }
\Address{\affiliation}
\vfill
\begin{Abstract}

Separate, validated implementations of the ATLAS and CMS new physics analyses are necessary to fully exploit the potential of these searches. To this end, we use \ma, a public framework for collider phenomenology.
In this talk, we present recent developments of \ma
, as well as a new public database of reimplemented LHC analyses.
The 
validation of one ATLAS and one CMS search for supersymmetry, present in the database, is also summarized.

\end{Abstract}
\vfill

\begin{Presented}
The Second Annual Conference\\
 on Large Hadron Collider Physics \\
Columbia University, New York, U.S.A \\ 
June 2-7, 2014
\end{Presented}
\vfill
\end{titlepage}
\def\thefootnote{\fnsymbol{footnote}}
\setcounter{footnote}{0}
%

\normalsize 


\section{Introduction}

The ATLAS and CMS collaborations have performed a large number of searches for new physics during Run~I of the LHC, targeting in particular supersymmetry in analyses based on missing transverse momentum. The implications of the (so far) negative results for new physics go well beyond the interpretations given in the experimental papers. Separate, validated implementations of the analyses using public fast simulation tools 
are necessary for theorists to fully exploit the potential of these searches. This will also give useful feedback to the experiments on the impact of their searches.

Recent developments of \ma~\cite{Conte:2014zja,Conte:2012fm}, the framework we use for reimplementing analyses, are presented in Section~\ref{sec:MA5new}.
The public database of reimplemented LHC analyses is then introduced in Section~\ref{sec:pad}.
Finally, a summary of the 
validation of one ATLAS and one CMS search for supersymmetry (SUSY) can be found in Sections~\ref{sec:atlas-susy-13-05} and~\ref{sec:cms-sus-13-011}, and conclusions are given in Section~\ref{sec:conclusions}.

\section{New developments in \ma}\label{sec:MA5new}

In most experimental analyses performed at the LHC, and in particular
the searches considered in this work, a branching set of
selection criteria (``cuts'') is used to define several
different sub-analyses (``regions'') within the same analysis.
In conventional coding frameworks, multiple regions are implemented with a nesting
of conditions checking these cuts, which grows exponentially more complicated
with the number of cuts. The scope of this project has therefore motivated us to
extend the \ma\ package to facilitate the handling of analyses with multiple regions,
as described in detail
in~\cite{Conte:2014zja}.

From version 1.1.10 onwards, the implementation of an analysis in the \ma\ framework
consists of implementing three basic functions: 
{\it i)}~\texttt{Initialize}, dedicated to the initialization of the signal regions,
    histograms, cuts and any user-defined variables;
{\it ii)}~\texttt{Execute}, containing the analysis cuts and weights applied to each event; and
{\it iii)}~\texttt{Finalize}, controlling the production of the results of the analysis, \ie,
    histograms and cut-flow charts.
To illustrate
the handling of multiple regions, we present a few
snippets of our implementation \cite{ma5code:cms-sus-13-011}
of the CMS search for stops in final states with one lepton~\cite{Chatrchyan:2013xna}
(see Section \ref{sec:cms-sus-13-011}).
This search comprises 16 signal regions (SRs), all of which must be declared in the
\texttt{Initialize} function.
This is done through the \texttt{AddRegionSelection} method
of the analysis manager class, of which \texttt{Manager()} is an instance provided
by default with each analysis. It takes as its argument
a string uniquely defining the SR under consideration.
For instance, two of the 16 SRs of the CMS analysis are declared as
\begin{verbatim}
 Manager()->AddRegionSelection("Stop->t+neutralino,LowDeltaM,MET>150");
 Manager()->AddRegionSelection("Stop->t+neutralino,LowDeltaM,MET>200");
\end{verbatim}

The \texttt{I\-ni\-ti\-a\-li\-ze}
function
should also contain the declaration of selection cuts. This
is handled by the \texttt{AddCut} method of
the analysis manager class. If a cut is common to all SRs, the
\texttt{AddCut} method takes as a single argument a string that uniquely identifies the cut.
An example of the declaration of two common cuts is
\begin{verbatim}
 Manager()->AddCut("1+ candidate lepton");
 Manager()->AddCut("1 signal lepton");
\end{verbatim}
If a cut is not common to all regions,
the \texttt{AddCut} method requires a second argument, either a
string or an array of strings, consisting of the names of all the regions to which
the cut applies. For example, an $E_T^{\rm miss}>150$~GeV cut that applies to four SRs
could be declared as
\begin{verbatim}
 string SRForMet150Cut[] = {"Stop->b+chargino,LowDeltaM,MET>150",
                            "Stop->b+chargino,HighDeltaM,MET>150",
                            "Stop->t+neutralino,LowDeltaM,MET>150",
                            "Stop->t+neutralino,HighDeltaM,MET>150"};
 Manager()->AddCut("MET>150GeV",SRForMet150Cut);
\end{verbatim}

Histograms are initialized in a similar fashion using the \texttt{AddHisto} method
of the manager class. A string argument is hence required
to act as a unique identifier for the histogram, provided together with its number
of bins and bounds. A further optional argument consisting
of a string or array of strings can then be used to associate it with specific
regions. The exact syntax can be found in the manual~\cite{Conte:2014zja}.

Most of the logic of the analysis is implemented in the \texttt{Execute} function.
This relies both on standard methods to declare particle objects and to compute
the observables of interest for event samples including detector
simulation~\cite{Conte:2012fm} and on the new manner in which cuts are
applied and histograms filled via the analysis manager class~\cite{Conte:2014zja}.
%
Below we provide a couple of examples for applying cuts and filling
histograms. After having declared and filled two vectors,
\texttt{Si\-gnal\-E\-lec\-trons} and \texttt{SignalMuons}, with objects satisfying the signal
lepton definitions used in the CMS-SUS-13-011 analysis,
we
require exactly one signal lepton with the following selection cut:
\begin{verbatim}
 if( !Manager()->ApplyCut( (SignalElectrons.size()+SignalMuons.size())>0,
   "1+ candidate lepton") ) return true;
\end{verbatim}
The \texttt{if(...)} syntax guarantees that a given event is discarded 
as soon as all regions fail the cuts applied so far.
Histogramming is as easy as applying a cut. For example, as we are interested in
the transverse-momentum distribution of the leading lepton, our code contains
\begin{verbatim}
 Manager()->FillHisto("pT(l)",SignalLeptons[0]->pt());
\end{verbatim}
This results in the filling of a histogram, previously declared with
the name \texttt{"pT(l)"} in the \texttt{Initialize} method, but only
when all cuts applied to the relevant regions are satisfied.


After the execution of the program, a set of {\sc Saf} files (an {\sc Xml}-inspired
format used by \ma) is created. They contain general information on the analyzed events,
as well as the cut-flow tables for all SRs and the histograms.
The structure of the various {\sc Saf} files is detailed in~\cite{Conte:2014zja}.

\section{Public analysis database of LHC new physics searches} \label{sec:pad}

A public database of reimplemented analyses in the \madanalysis\ framework and using \delphes~3~\cite{deFavereau:2013fsa} was presented in~\cite{Dumont:2014tja}.
The list of analyses presently available in the database can be found on the wiki page~\cite{ma5wiki}. Each analysis code, in the \cpp\ language used in \madanalysis, is submitted to INSPIRE
, hence is searchable and citeable.
The information on the number of background and observed events is required for setting limits and is provided in the form of an {\sc Xml} file that is submitted to INSPIRE together with the analysis code.
Finally, detector tunings (contained in the detector card for \delphes) as well as detailed validation results for each analysis can be found on the wiki page.
To date, there are five SUSY analyses in the database, two from ATLAS and three from CMS.

From an event file in \texttt{StdHep} or \texttt{HepMc} format, the acceptance$\times$efficiency can be found in the output of \ma\ for each SR.
The limit setting can subsequently done under the CL$_s$ prescription with the code {\tt exclusion\_CLs.py}. It reads the cross section and the acceptance$\times$efficiency from the output of \madanalysis, while the luminosity and the required information on the signal regions is taken from the {\sc Xml} file mentioned above. 


\section{ATLAS-SUSY-2013-05: search for third-generation squarks 
in final states with zero leptons and two $b$-jets}
 \label{sec:atlas-susy-13-05}

In this ATLAS analysis \cite{Aad:2013ija}, stops and sbottoms 
are searched for in final states with 
large missing transverse momentum and two jets identified as $b$-jets. The
results are presented for an integrated luminosity of $20.1$~fb$^{-1}$ at
$\sqrt{s} = 8$ TeV. Two possible sets of SUSY mass spectra were
investigated in this analysis:
{\it i)}~sbottom $\tilde{b}_1$ pair production with $\tilde{b}_1 \rightarrow b\tilde{\chi}_1^0$, and
{\it ii)}~stop $\tilde{t}_1$ pair production with $\tilde{t}_1 \rightarrow b
\tilde{\chi}_1^\pm$, where the subsequent decay of the $\tilde{\chi}_1^\pm$ is
 invisible due to a small mass splitting with the $\tilde{\chi}_1^0$.
Two sets of SRs, SRA and SRB, are defined to provide sensitivity
to respectively large and small mass splittings between the squark and the neutralino.

\begin{table*}[!t]
\begin{center}
\begin{tabular}{ l ||c|c||c|c}
& \multicolumn{2}{|c||}{$m_{\tilde b_1}=350$~GeV} &
\multicolumn{2}{c}{$m_{\tilde t_1}=500$~GeV}  \\
cut & ATLAS result & {\sc MA}\,5 result & ATLAS result & {\sc MA}5 result \\ 
\hline\noalign{\smallskip} 
$E^{\rm miss}_T> 80$~GeV~filter  & $6221.0$  & $5963.7$ & $1329.0$ & $1117.9$\\ 
+ Lepton veto & $4069.0$ & $4987.9$  & $669.0$ & $932.9$  \\
+ $E^{\rm miss}_T > 250$~GeV & $757.0$  & $802.9$ & $93.0$ & $117.2$  \\
+ Jet Selection & $7.9$ & $5.4$ & $6.2$ & $5.3$ \\ 
+ $H_{T,3} < 50$~GeV & $5.2$  & $4.6$ & $3.0$ & $4.2$\\ 
\noalign{\smallskip}\hline
\end{tabular}
\end{center}
\caption{Summary of yields for SRB of ATLAS-SUSY-2013-05 corresponding to the benchmark points 
$(m_{\tilde b_1}, m_{\tilde \chi^0_1}) = (350,320)$~GeV and 
$(m_{\tilde t_1},m_{\tilde \chi^\pm_1}, m_{\tilde\chi^0_1})=(500,420,400)$~GeV,
as compared to official ATLAS results given on \cite{Aad:2013ija}. 
An $E^{\rm miss}_T$ filter is applied at the particle level. See \cite{Aad:2013ija} for more detail. 
\label{tab:atlas-13-05-cutflow-SRB}}
\end{table*}

The analysis is very well documented regarding physics, but for
recasting purposes more information than provided in~\cite{Aad:2013ija} 
was needed. 
This made the validation of the recast code seriously difficult 
in the earlier stages of the project. 
Since then, fortunately, cut-flow tables were made public, as well as SUSY Les Houches Accord (SLHA) input files and the exact version of Monte Carlo tools used to generate the signal.
However, the collaboration did not provide information on trigger-only and $b$-tagging efficiencies.

\begin{figure*}[!t]
\centering
\includegraphics[width=5.2cm]{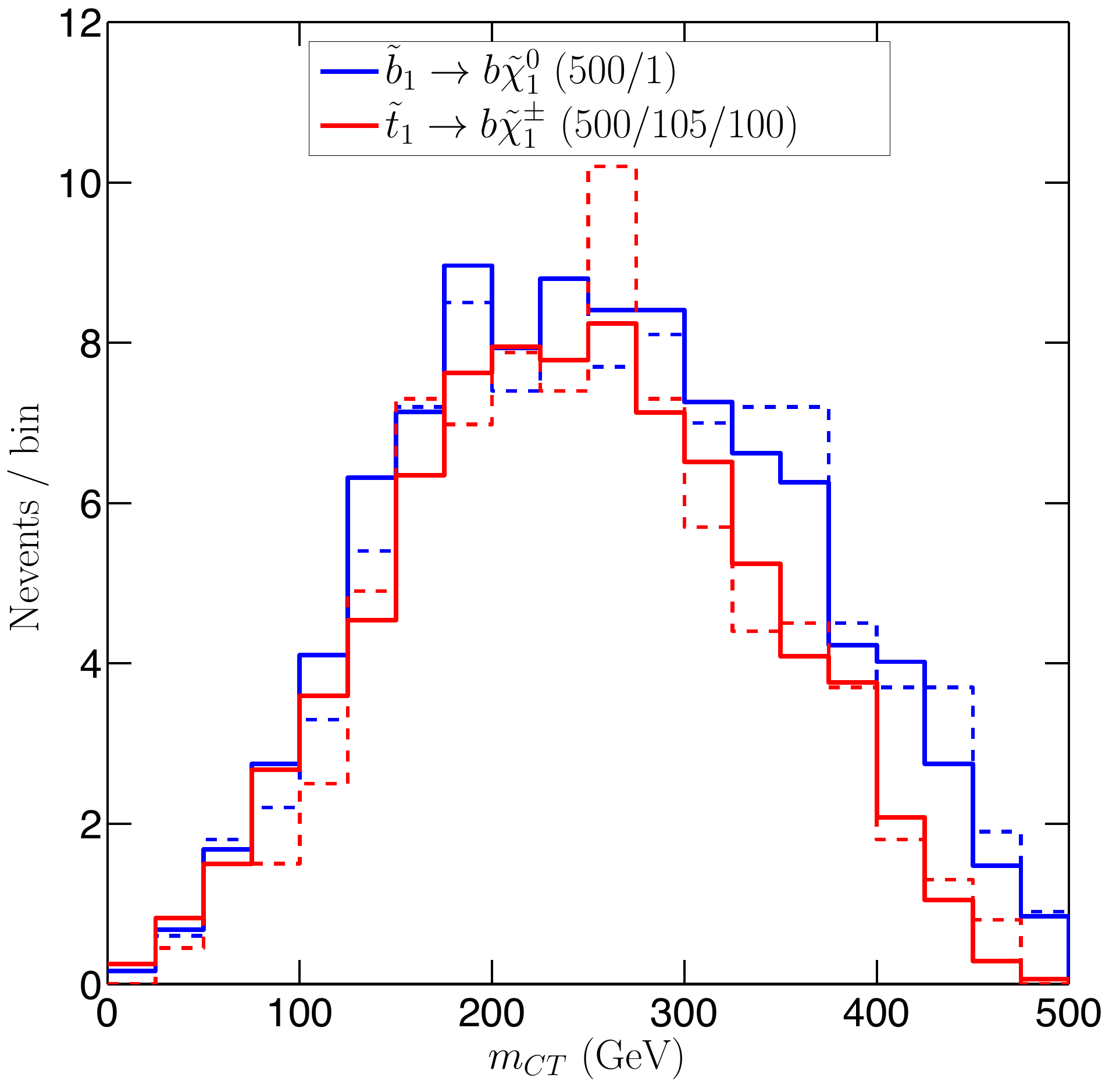} \qquad
\includegraphics[width=5.2cm]{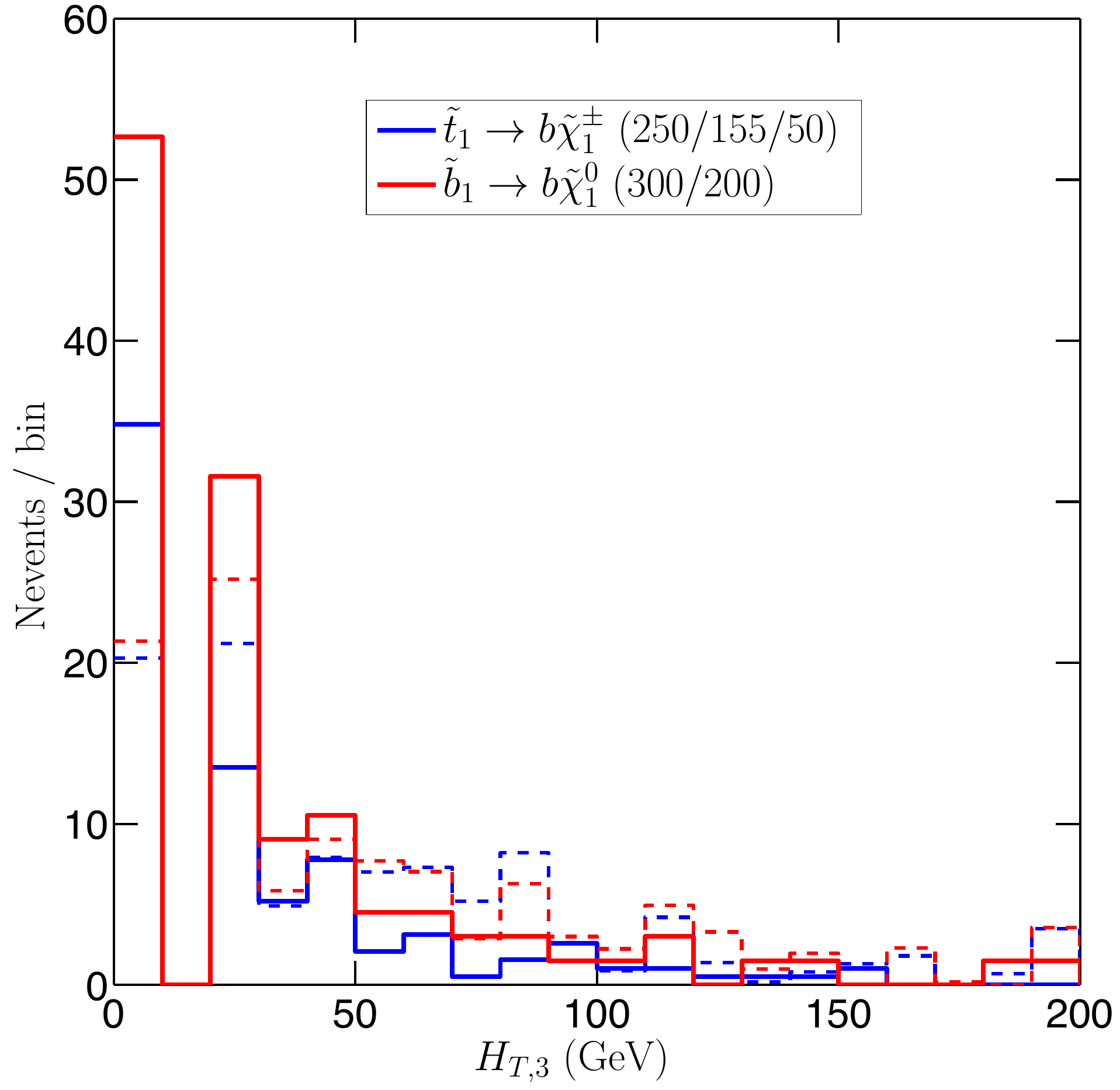}
\caption{\label{fig:SRAHistos}Distributions of $m_{CT}$ for SRA and $H_{T,3}$ for SRB of ATLAS-SUSY-2013-05 without their respective cut. On the left plot, the benchmark points used are $(m_{\tilde b_1},m_{\tilde\chi_1^0})$ = $(500,1)$~GeV (in blue) 
and $(m_{\tilde t_1},m_{\tilde \chi^\pm_1},m_{\tilde \chi^0_1})$ = $(500,105,100)$~GeV (in red). On the right plot, $(m_{\tilde t_1},m_{\tilde \chi^\pm_1},m_{\tilde \chi^0_1})$ = $(250,155,50)$~GeV (in blue) and $(m_{\tilde b_1},m_{\tilde\chi_1^0})$ = $(300,200)$~GeV (in red).
The solid lines correspond to our re-interpretation within \ma\ and the dashed lines to the ATLAS result.}
\end{figure*}


The comparison between the official cut flows and the ones obtained within \ma\ is presented 
in the case of SRB in Table~\ref{tab:atlas-13-05-cutflow-SRB}. Moreover, distributions of the contransverse variable $m_{CT}$ and of $H_{T,3}$ are shown in Fig.~\ref{fig:SRAHistos}. ($H_{T,3}$ is defined as the sum of the $p_T$ of the $n$ jets without including the leading three jets.)
%
The largest discrepancy is observed in SRB, as be seen in the distribution of $H_{T,3}$.
To investigate this issue more deeply, a
more detailed cut flow about the ``Jet selection'' line in Table~\ref{tab:atlas-13-05-cutflow-SRB} 
would be appreciable since it directly
impacts the $H_{T,3}$ variable.

Overall the agreement is quite satisfactory, considering the expected accuracy
for a fast simulation.
For SRA the agreement is very good. For SRB, the importance of the treatment of soft jets induces a sizable discrepancy with respect to the ATLAS results. Further tunings of the fast detector simulation are needed, and are currently under investigation.
However, the current results (for which detailed validation material can be found at~\cite{ma5wiki}) lead us to conclude that this implementation is validated.
The \ma\ recast code is available as~\cite{ma5code:atlas-susy-2013-05}.

\section{CMS-SUS-13-011: search for stops in the single-lepton final state} \label{sec:cms-sus-13-011}

The CMS search for stops in the single lepton and missing energy, $\ell + E^{\rm miss}_T$, final state with full luminosity at
$\sqrt{s} = 8$~TeV~\cite{Chatrchyan:2013xna} has been taken as a ``template analysis'' to develop a common language and framework for the analysis implementation. 
The analysis targets two possible decay modes of the stop: $\tilde{t} \to t \tilde{\chi}^{0}_1$ and 
$\tilde{t} \to b \tilde{\chi}^{+}_1$.  
Since the stops are pair-produced, their decays give rise to two $W$ bosons in each event, one of which is assumed to  decay leptonically, while the other one is assumed to decay hadronically. 
In the cut-based version of the analysis, 
 that we consider, 
 two sets of signal regions with different cuts, each  dedicated to one of the two decay modes, are defined. These two sets are further divided into ``low $\Delta M$'' and ``high $\Delta M$'' categories, targeting small and large mass differences with the lightest neutralino $\tilde\chi_1^0$, respectively. Finally, each of these four categories are further sub-divided using four different $E^{\rm miss}_T$ requirements. In total, 16 different, potentially overlapping SRs are defined. 

Overall, this analysis is well documented.
Detailed trigger efficiencies and the identification-only efficiencies for electron and muons were provided by the CMS collaboration upon request and are now available on the analysis Twiki page~\cite{Chatrchyan:2013xna} in the section ``Additional Material to aid the Phenomenology Community with Reinterpretations of these Results''.
The $b$-tagging efficiency as a function of $p_T$ was taken from~\cite{Chatrchyan:2013fea}.
Another technical difficulty came from the isolation criteria.
Since we used a simplified isolation criteria, we applied on the events a weighting factor of $0.885$ that was determined from the two cut flows (see Table~\ref{tab:cms-13-011-cutflow}).
%

\begin{figure*}[!t]\centering
\includegraphics[width=5.5cm]{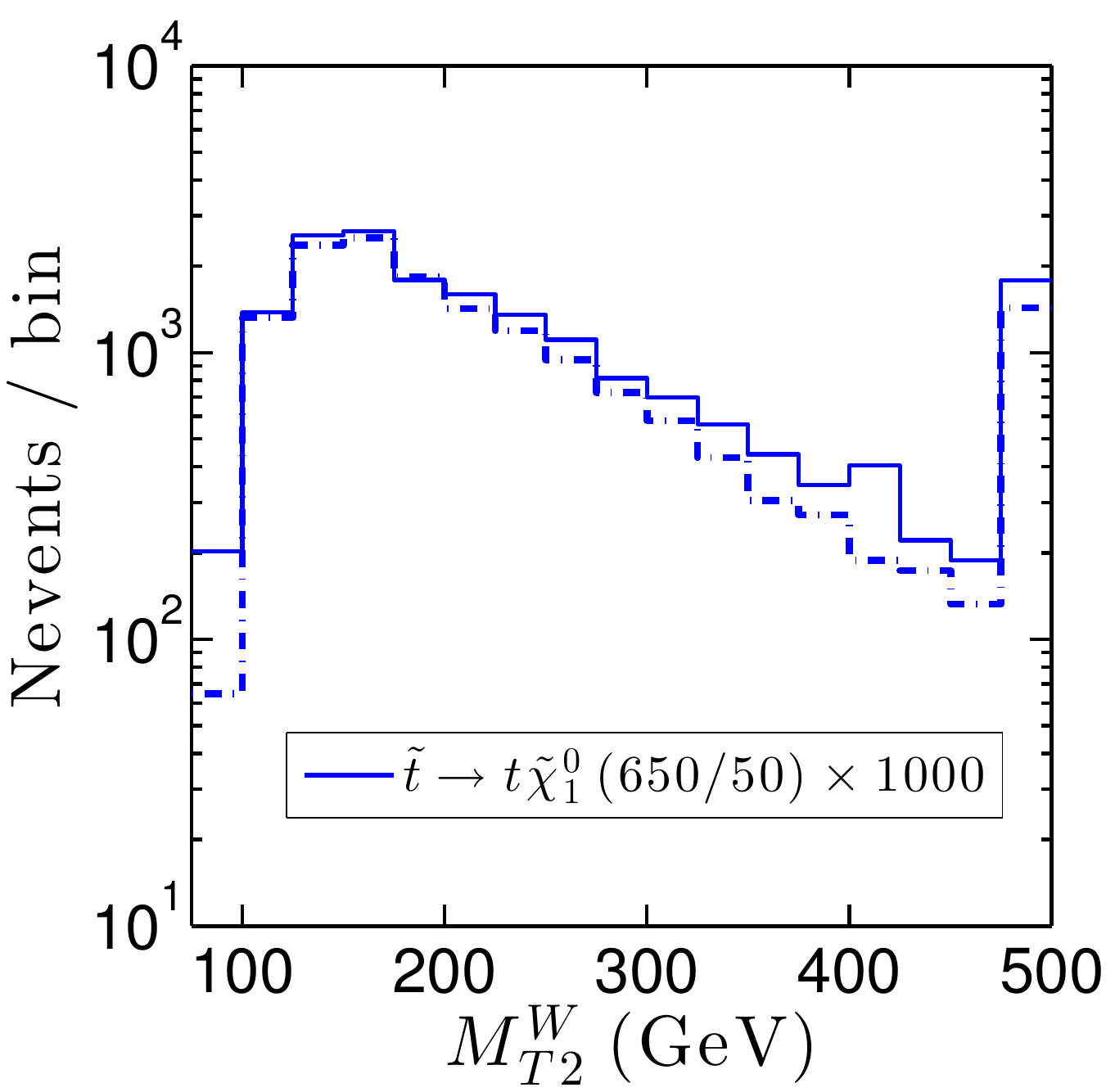}\quad
\includegraphics[width=5.5cm]{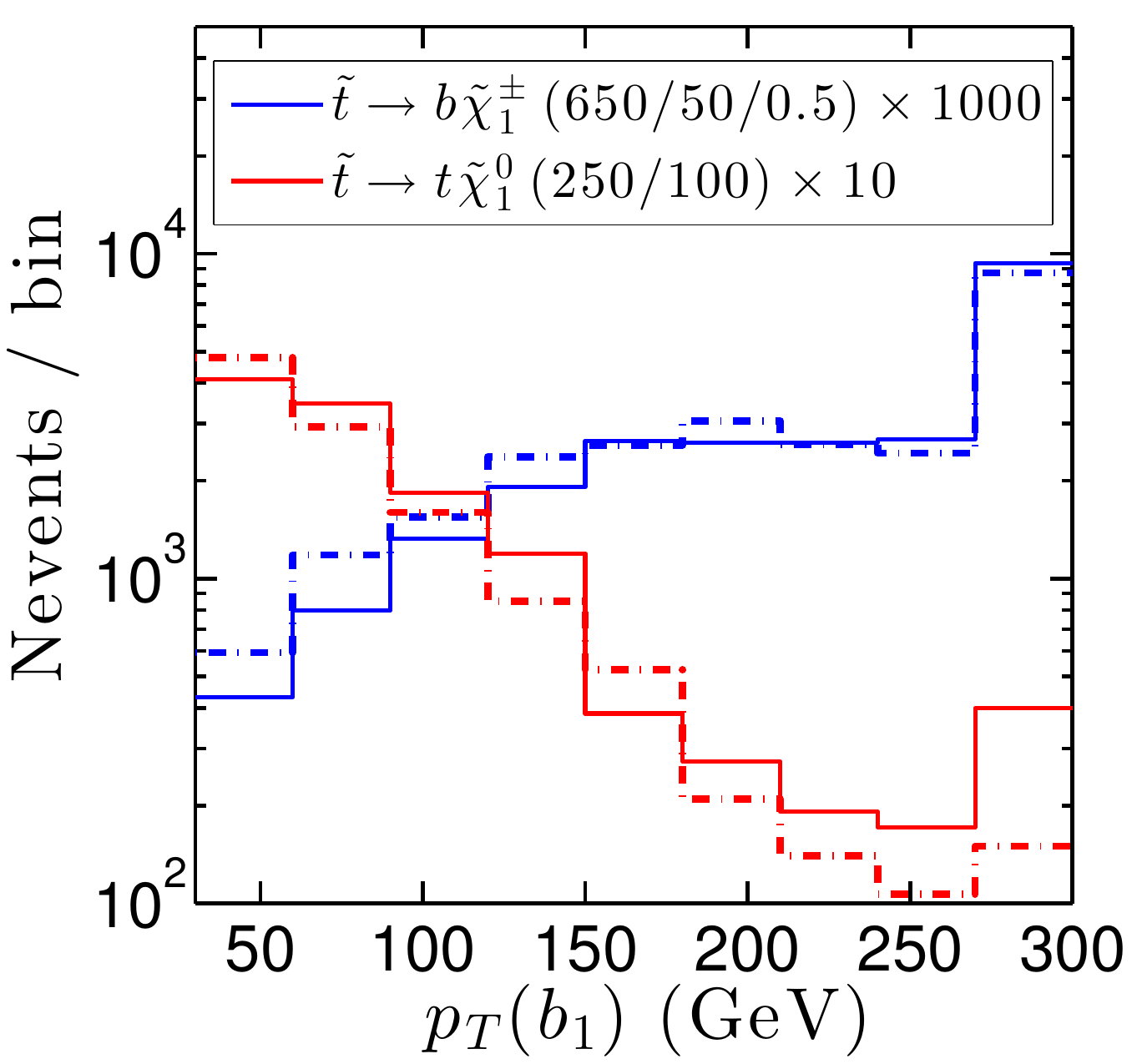}
\caption{Distributions of $M^W_{T2}$ (left) and of the $p_T$ of the leading $b$-tagged jet (right) after the preselection cuts of the analysis CMS-SUS-13-011. The solid lines are obtained from our re-interpretation within \ma, while the dash-dotted lines correspond to the CMS results, given in Fig.~2 of~\cite{Chatrchyan:2013xna}. 
} \label{fig:kinvarsus13011}
\end{figure*}

%
The validation was done using the eleven benchmark points 
presented in the experimental paper. 
%
The validation process was based on (partonic) event samples, in LHE format, provided by the CMS collaboration. 
%
Some examples of histograms reproduced for the validation are shown in Fig.~\ref{fig:kinvarsus13011}. The shapes of the distributions shown---as well as all other distributions that we obtained but do not show here---follow closely the ones from CMS, which indicates the correct implementation of the analysis and all the kinematic variables. 

Upon our request, the CMS SUSY group furthermore provided detailed cut-flow tables, which are now also available on the Twiki page of the analysis~\cite{Chatrchyan:2013xna}. 
These proved extremely useful because they allowed us to verify our implementation step-by-step in the analysis. 
A comparison of our results with the official CMS ones is given in Table~\ref{tab:cms-13-011-cutflow}. 
For both cases shown, CMS results are reproduced within about 20\%.  
On the whole, we conclude that our implementation gives reasonably accurate results 
(to the level that can be expected from fast simulation). 
The \ma\ code for this analysis, including extensive comments, is published as \cite{ma5code:cms-sus-13-011}.

\begin{table*}[!t]
\begin{center}
\begin{tabular}{ l ||c|c||c|c}
& \multicolumn{2}{|c||}{$m_{\tilde t}=650$~GeV} & \multicolumn{2}{c}{$m_{\tilde t}=250$~GeV}  \\
cut & CMS result & {\sc MA}\,5 result & CMS result & {\sc MA}5 result \\ 
\hline\noalign{\smallskip}
$1\ell\, + \ge 4{\rm jets} + E_T^{\rm miss}>50$~GeV & $31.6\pm0.3$ & $29.0$ & $8033.0\pm38.7$ &  $7365.0$  \\ 
+ $E_T^{\rm miss}>100$~GeV   & $29.7\pm0.3$ & $27.3$ & $4059.2\pm 27.5$  & $3787.2$ \\
+ $n_b\ge1$        & $25.2\pm0.2$ & $23.8$ & $3380.1\pm25.1$ & $3166.0$ \\
+ iso-track veto   & $21.0\pm0.2$ & $19.8$ & $2770.0\pm22.7$ & $2601.4$ \\
+ tau veto             & $20.6\pm0.2$ & $19.4$ & $2683.1\pm22.4$ & $2557.2$ \\
+ $\Delta\phi_{\rm min}>0.8$  & $17.8\pm0.2$ & $16.7$ & $2019.1\pm19.4$ & $2021.3$ \\
+ hadronic $\chi^2<5$  & $11.9\pm0.2$ & $9.8$ & $1375.9\pm16.0$ & $1092.0$ \\
+  $M_T>120$~GeV & $9.6\pm0.1$ & $7.9$ & $355.1\pm8.1$ & $261.3$ \\
${\rm high\,} \Delta M, E^{\rm miss}_T > 300~{\rm GeV}$ & $4.2\pm0.1$ & $3.9$ & --- & ---\\
${\rm low\,} \Delta M, E^{\rm miss}_T > 150~{\rm GeV}$ & --- & --- & $124.0\pm4.8$ & $107.9$\\
\noalign{\smallskip}\hline
\end{tabular}
\end{center}
\caption{Summary of yields for the $\tilde{t} \rightarrow t \tilde{\chi}^0_1$ model for two benchmark points with 
$m_{\tilde\chi^0_1}=50$~GeV, as compared to official CMS-SUS-13-011 results given on~\cite{Chatrchyan:2013xna}. 
The uncertainties given for the CMS event numbers are statistical only.  
See \cite{Chatrchyan:2013xna} for more details on the definition of the cuts.  
\label{tab:cms-13-011-cutflow}}
\end{table*}

\section{Conclusions} \label{sec:conclusions}

We presented recent developments of \ma\ that were necessary for the implementation and for recasting LHC new physics analyses. After validation, reimplemented analyses are stored in a new public database.
We discussed the validation of two SUSY analyses. A growing number of such analysis codes, including detailed validation material, is being made available in a public analysis database, see~\cite{ma5wiki}.

\Acknowledgements
I am grateful to the ATLAS and CMS collaborations for their help in validating the results. I would also like to thank my colleagues of~\cite{Conte:2014zja} and~\cite{Dumont:2014tja} for their collaboration and for inspiring discussions.


\begin{thebibliography}{99}


\bibitem{Conte:2014zja}
  E.~Conte, B.~Dumont, B.~Fuks and C.~Wymant,
  arXiv:1405.3982 [hep-ph].

\bibitem{Conte:2012fm}
  E.~Conte, B.~Fuks and G.~Serret,
  Comput.\ Phys.\ Commun.\  {\bf 184} (2013) 222
  [arXiv:1206.1599 [hep-ph]].

\bibitem{ma5code:cms-sus-13-011}
B.~Dumont, B.~Fuks, and C.~Wymant,
doi:~\href{http://doi.org/10.7484/INSPIREHEP.DATA.LR5T.2RR3}{10.7484/INSPIREHEP.DATA.LR5T.2RR3}.  

\bibitem{Chatrchyan:2013xna}
  S.~Chatrchyan {\it et al.}  [CMS Collaboration],
  Eur.\ Phys.\ J.\ C {\bf 73} (2013) 2677
  [arXiv:1308.1586 [hep-ex]];
  \url{https://twiki.cern.ch/twiki/bin/view/CMSPublic/PhysicsResultsSUS13011}.

\bibitem{deFavereau:2013fsa}
  J.~de Favereau {\it et al.}  [DELPHES 3 Collaboration],
  JHEP {\bf 1402} (2014) 057
  [arXiv:1307.6346 [hep-ex]].

\bibitem{Dumont:2014tja}
  B.~Dumont, B.~Fuks, S.~Kraml, {\it et al.},
  arXiv:1407.3278 [hep-ph].

\bibitem{ma5wiki}
 \url{http://madanalysis.irmp.ucl.ac.be/wiki/PhysicsAnalysisDatabase}.

\bibitem{Aad:2013ija}
  G.~Aad {\it et al.}  [ATLAS Collaboration],
  JHEP {\bf 1310} (2013) 189
  [arXiv:1308.2631 [hep-ex]]; \\
  \url{https://atlas.web.cern.ch/Atlas/GROUPS/PHYSICS/PAPERS/SUSY-2013-05/}.

\bibitem{ma5code:atlas-susy-2013-05}
G.~Chalons,
doi:~\href{http://doi.org/10.7484/INSPIREHEP.DATA.Z4ML.3W67}{10.7484/INSPIREHEP.DATA.Z4ML.3W67}.

\bibitem{Chatrchyan:2013fea}
  S.~Chatrchyan {\it et al.}  [CMS Collaboration],
  JHEP {\bf 1401} (2014) 163
  [arXiv:1311.6736 [hep-ex]].

\end{thebibliography}
\end{document}